\documentclass[12pt]{article}	
\usepackage[T1]{fontenc}		
\usepackage[utf8]{inputenc}		
\usepackage{amsmath}
\usepackage{amsfonts}
\usepackage{amssymb}
\usepackage{indentfirst}
\usepackage{color}
\usepackage{array}
\usepackage{esvect}

\usepackage{times, graphicx, graphics, enumerate, hyperref, color, bm}

\begin{document}

\title{
Circular Dichroism without absorption in
isolated chiral dielectric Mie particles} 

\author{ (1) Rafael S. Dutra\footnote{e-mail: rafael.dutra@ifrj.edu.br } , (2) Felipe A. Pinheiro, (2) Diney S. Ether Jr, \\ (3) Cyriaque Genet, (2) Nathan B. Viana,  (2) Paulo A. Maia Neto \footnote{e-mail: pamn@if.ufrj.br}
\\(1) Laboratório de Instrumentação e Simulação Computacional,\\ Instituto Federal do Rio de Janeiro, 26600-000, Paracambi-RJ, Brazil
\\(2) Instituto de F\'isica, Universidade Federal do Rio de Janeiro, \\ Caixa Postal 68528, Rio de Janeiro, Rio de Janeiro 21941-972, Brazil
\\(3) Université de Strasbourg, CNRS,\\ Institut de Science et d’Ingénierie Supramoléculaires, \\UMR 7006, F- 67000 Strasbourg, France}


\date{\today}
\maketitle
\begin{abstract}

We demonstrate that an effect phenomenologically analogous to circular dichroism can arise even for dielectric and isotropic chiral spherical particles. By analyzing the polarimetry of light scattered from a chiral, lossless microsphere illuminated with linearly polarized light, we show that the scattered light becomes nearly circularly polarized, exhibiting large, nonresonant values of the Stokes parameter $S_3$ for a broad range of visible frequencies. This phenomenon occurs only in the Mie regime, with the microsphere radius comparable to the wavelength, and provided that the scattered light is collected by a high-NA objective lens, including non-paraxial Fourier components. Altogether, our findings offer a theoretical framework and motivation for an experimental demonstration of a novel chiroptical effect with isolated dielectric particles, with potential applications in enantioselection and characterization of single microparticles, each and every one with its own chiral response.   

\end{abstract}

\section{Introduction}

An object is considered chiral if it has non-superposable mirror images, {\it i.e. two enantiometers}. The separation of chiral enantiomers is a significant scientific and technological challenge with broad, multidisciplinary applications~\cite{barron04,wagniere2007chirality,schaferling17}. Chirality also shows up in the optical properties of materials in a very characteristic way, so that the chiroptical response provides one of the most direct and effective means for analyzing chiral systems. Indeed, chiral objects exhibit differential absorption of left- and right-handed circularly polarized light, known as Circular Dichroism (CD)~\cite{barron04,wagniere2007chirality,schaferling17}. Additionally, chiral systems can rotate the plane of incident linearly polarized light in a direction determined by their handedness, a phenomenon known as optical rotatory power~\cite{barron04,wagniere2007chirality,schaferling17}.

CD spectroscopy is one of the most traditional methods employed for the enantioselective detection of chiral molecules \cite{pignataro20}. The resulting CD spectra are unique to a molecule’s specific conformation, with the sign of the signal indicating the enantiomer's handedness. However, the intrinsically weak chiroptical signals fundamentally limit the sensitivity of CD spectroscopy so that it typically probes bulk samples. The advent of nanophotonics and plasmonics has led to the development of various strategies to enhance CD, thereby enabling more efficient chiral sensing~\cite{warning2021nanophotonic,lininger22}. The limitations imposed by intrinsically weak CD signals become particularly pronounced in single-molecule studies sensing, requiring special techniques in the case of individual molecules, for instance fluorescence-detected circular dichroism~\cite{hassey06}, or nonlinear resonant phasesensitive microwave spectroscopy~\cite{patterson13}, as well as in the characterization of larger, isolated chiral microparticles,
which are promising platforms for applications in nanophotonics~\cite{zerrouki08,hernandez13,hernandez16,huang15,huang18,meng20}, such as enantioselection via optical forces~\cite{genet2022chiral,yang25,man24,ali2020enantioselective,ali2021enantioselection,ali2023enantioselective,ali2024enantioselective,Diniz2025}. 
To circumvent these limitations, different strategies that include substrate-assisted CD~\cite{nechayev2019substrate}, extrinsic chirality~\cite{lu2014circular}, imaging techniques~\cite{huang21,narushima2016circular}, and plasmonic materials have been employed to enhance the weak chiroptical response of single chiral nanoparticles~\cite{lee24,adhikari2022optically,lu2014circular,zhang2019unraveling}. 
Indeed, thanks to the strong interaction between light and free electrons chiral plasmonic nanoparticles exhibit distinctive resonances that enables the experimental observation of single-particle CD spectroscopy~\cite{schnoering2018three,adhikari2022optically,vinegrad2018circular}. This technique allows for the detection of CD in individual chiral nanoparticles by measuring differences in extinction, scattering, or absorption between left- and right-circularly polarized light enabling enantiomeric recognition~\cite{schnoering2018three,adhikari2022optically}. In addition to designing plasmonic particles with chiral geometries~\cite{karst19,fan2012chiral}, other strategies to enhance chiroptical effects include synthesizing plasmonic systems in the presence of chiral molecules or under conditions breaking mirror symmetry \cite{kant24}, chiral optical cavities, and photothermal approaches ~\cite{im24}. By enhancing the chiral optical response, these strategies allow for enantioselection and chiral characterization at the scale of single nanoparticles. However, since these approaches typically rely on the excitation of plasmonic resonances, the enhancement of chiroptical properties in single nanoparticles often comes at the expense of high losses and limited frequency bandwidths~\cite{hentschel17,garcia2019enhanced,sun23}.

In this context, the development of alternative mechanisms for probing the chiroptical response of individual chiral nanoparticles, without relying on plasmonic effects, has become increasingly important. This need is further underscored by the recent developments and applications of all-dielectric chiral nanosystems, such as optical cavities for enhanced chiral sensing~\cite{khanbekyan2022enantiomer,feis2020helicity}. These cavities can be Mie particles that support high quality factor resonances~\cite{olmos2022helicity,mohammadi2023nanophotonic,olmos2024capturing}, tailored  to assist and facilitate chiral sensing, chiral transfer and enantioselection~\cite{olmos2022helicity,mohammadi2023nanophotonic,olmos2024capturing}.

Building on these motivations, the present study reveals a novel chiroptical response in single, lossless, and isotropic chiral Mie microspheres, which phenomenologically manifests as the well-known circular dichroism (CD) observed in absorbing media. Specifically, we demonstrate that linearly polarized light scattered by such particles—and collected using a high-numerical-aperture (NA) objective lens—becomes nearly circularly polarized, leading to enhanced, nonresonant values of the Stokes parameter $S_3$ across a broad range of visible frequencies, in contrast to cases assisted by plasmonic resonances. We show that this effect arises intrinsically from the non-paraxial Fourier components of the scattered light and the underlying Mie scattering regime. These findings not only reveal a CD-like response in lossless particles but also open new avenues for applying Mie-tronics to chiral sensing, chiral characterization, and enantioselective technologies~\cite{kivshar22}.

\section{Theoretical Model}

\begin{figure}[t!]
    \centering
 \includegraphics[scale=0.6]{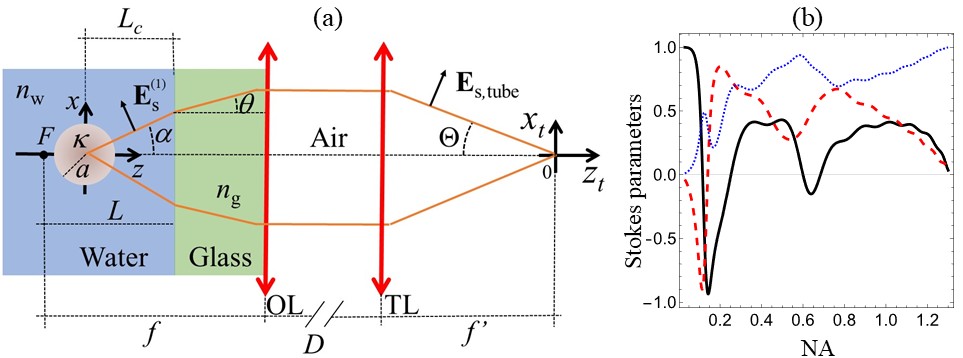}
 \vspace{-0.3cm}
   \caption{(a) Imaging configuration in an optical microscope with the collection of forward-scattered illumination. The incident illumination, described by a plane wave, is scattered by the microsphere, with radius $a$ and chirality parameter $\kappa$. Then, the illumination is collected by the objective lens OL and finally focused by the tube lens TL with respective focal lengths $f$ and $f'$. (b) Stokes parameters $S_1$ (black solid line), $S_2$ (red dashed line), and $S_3$ (blue dotted line), normalized by the Stokes parameter $S_0$, as functions of the numerical aperture (NA) of the objective lens, for 
   a chirality parameter set to 
   $\kappa=-0.02$ and a wavelength
    $\lambda=0.464\,\mu{\rm m}.$}

  \label{fig1} 
  \end{figure}

In this section we describe the model to achieve imaging of chiral, homogeneous Mie microspheres in the forward direction. The incident illumination on the microsphere is described as a plane wave of wavelength $\lambda$ propagating in water with a wave vector magnitude $k_{\rm w}=2\pi n_{\rm w}/\lambda$ and linearly polarized along the $\hat{\mathbf{x}}$ direction, represented by the electric field
\begin{equation}\label{Ein}
\mathbf{E}=E_0e^{i(k_{\rm w}\,z-\omega t)}\,\hat{\mathbf{x}},
\end{equation}
where $n_{\rm w}$ is the refractive index of water. The setup is schematically depicted in Fig. \ref{fig1}(a).

We assume that the microsphere is composed of a homogeneous and isotropic chiral material in which the electric and magnetic fields are coupled according to the following constitutive relations \cite{sihvola91}:

\begin{equation}
  \begin{pmatrix}
  \mathbf{D} \\ \mathbf{B} 
  \end{pmatrix} = 
  \begin{pmatrix}
      \epsilon_0 \epsilon  & i \kappa/c \\
      - i\kappa/c & \mu_0 \mu
  \end{pmatrix}
  \begin{pmatrix}
  \mathbf{E} \\ \mathbf{H}
  \end{pmatrix},
  \label{eq:constitutive_equation}
\end{equation}
where $\epsilon$ and $\mu$ are the relative permittivity and permeability, $c = 1/\sqrt{\epsilon_0 \mu_0}$ is the speed of light in vacuum, and $\kappa$ is the pseudoscalar defined as the chirality parameter or chiral index of the medium (Pasteur parameter), which couples the electric and magnetic fields.

We expand the incident electric field (\ref{Ein}) as well as the corresponding magnetic field as superpositions of spherical multipole waves written in terms of Debye potentials\cite{bouwkamp54,panofsky12}. 
In the circular polarization basis, the scattering matrix of the chiral microsphere is diagonal in the representation defined by the electric (E) and magnetic (M) multipoles. Hence we expand the incident polarization vector $\hat{\mathbf{x}}=(\mathbf{\hat{\varepsilon}}^{+}+\mathbf{\hat{\varepsilon}}^{-})/\sqrt{2}$ in the circular polarization basis $\mathbf{\hat{\varepsilon}}^{\sigma}= (\mathbf{\hat{x}}+i \sigma \mathbf{\hat{y}})/\sqrt{2}$, with $\sigma=\pm 1$
 denoting the helicity\cite{poulikakos19}, and then 
 solve for the scattered field for each helicity component. 
 The scattered Debye potentials for helicity $\sigma$ associated with electric  and magnetic multipoles are  given by \cite{mazolli2003theory,ali2020probing, ali2021enantioselection}
\begin{equation}
\Pi^{E}_{s,\sigma}(r,\Theta,\Phi)= -\frac{\sigma E_{0}}{ k_{\rm w}}\sum_{j} (i)^{j+1}A_{j}^{\sigma}\sqrt{ \frac{4\pi(2j+1)}{j(j+1)} }Y_{j}^{\sigma}(\Theta,\Phi)h_{j}^{(1)}(k_{\rm w} r) \label{DE_s}
\end{equation}
and 
\begin{equation}
\Pi^{M}_{s,\sigma}(r,\Theta,\Phi)= -\frac{H_{0}}{k_{\rm w}}\sum_{j,m} (i)^jB_{j}^{\sigma}\sqrt{ \frac{4\pi(2j+1)}{j(j+1)} }Y_{j}^{\sigma}(\Theta,\Phi)h_{j}^{(1)}(k_{\rm w}r), \label{DH_s}
\end{equation}
where $Y_{j}^{\sigma}$ are the spherical harmonics.
Physically, the Hankel functions $h_{j}^{(1)}(k_{\rm w}r)$ describe the outgoing behavior of the scattered spherical waves. The expressions for the effective Mie scattering coefficients $A_j^{\sigma}$ and $B_j^{\sigma}$ for a chiral sphere of radius $a$ are presented in Appendix A. 
To describe light propagation of the scattered field through the optical system, we expand the scattered spherical waves into plane waves employing the integral Weyl representation \cite{devaney1974multipole, weyl1919ausbreitung, bobbert1986light, dutra2016theory}

\begin{eqnarray}
Y_{j}^{\sigma}(\Theta,\Phi)h_{j}^{(1)}(k_{\rm w}r)=  \frac{i^{-j}}{2\pi}\int_{0}^{2\pi}d\beta\int_C d\alpha\sin\alpha\,  
Y_{j}^{\sigma}(\alpha,\beta)e^{i\mathbf{k}_{\rm w}(\alpha,\beta)\cdot\mathbf{r}}. \label{weyl}
\end{eqnarray}
The direction of the wavevector \( \mathbf{k}_{\rm w}(\alpha,\beta)\) is determined by the spherical angles \( (\alpha,\beta).\)
The integration contour \(C\) is selected~\cite{bobbert1986light} to account for both evanescent waves (imaginary values of \(\alpha\)) and homogeneous waves that propagate in the forward direction  \(z > 0\) (\(k_{\rm w z} > 0\)). As a result, the scattered field expanded into plane waves in the aqueous solution  reads

\begin{eqnarray}\label{Es}
\mathbf{E}^{(\rm w)}_{\rm s}({\bf r})=-\frac{E_{0}}{8\pi}\sum_{\sigma=-1,+1}\sum_{j=1}^{\infty}(2j+1)
\int_{0}^{2\pi} d\beta\int_Cd\alpha&\sin\alpha\left[A_j^{\sigma}g^{j}_{\sigma,+}(\alpha,\beta)+B_j^{\sigma}g^{j}_{\sigma,-}(\alpha,\beta)\right]
\\
&\times
(\hat{\boldsymbol{\vartheta}}+ i\sigma\hat{\boldsymbol{\varphi}})
\,e^{i\mathbf{k}_{\rm w}(\alpha,\beta)\cdot\mathbf{r}}. \nonumber\end{eqnarray}
We have defined the coefficients 
\begin{equation}
g^{j}_{\sigma,\epsilon}(\alpha,\beta)  = e^{i\sigma\beta}d^{j}_{\sigma,\sigma}(\alpha) - \epsilon \,e^{-i\sigma\beta}d^{j}_{-\sigma,\sigma}(\alpha),
\end{equation}
where \(\epsilon = 1\) for electric multipoles and \(\epsilon = -1\) for magnetic multipoles. They are defined in terms of the matrix elements of finite rotations \(d^j_{m',m}(\alpha)\)~\cite{edmonds1996angular} with
\(m = m' = \sigma\) for terms that conserve helicity, and \(m = -m' = \sigma\) for the contributions due to spin-orbit helicity flip~\cite{bliokh2015spin,schwartz2006conservation} upon Mie scattering.

 The imaging setup consists of an optical microscope where the scattered light is initially collected by an oil immersion objective with numerical aperture ${\rm NA>1,}$ focal length $f$, and aperture angle $\theta_0=\arcsin({\rm NA}/n_g)$, where $n_g$ is the refractive index of the glass, as depicted in Fig.~\ref{fig1}(a). 
 Each Fourier component of the scattered field $\mathbf{E}^{(\rm w)}_{\rm s}({\bf r})$
given by (\ref{Es}) is 
 characterized by its wave vector $\mathbf{k}_{\rm w}(\alpha,\beta)$. 
 As the Fourier components propagate away from the microsphere, they first refract at the 
  interface between the sample chamber and the glass slide as shown in Fig.~\ref{fig1}(a).
In addition to a reduction of amplitude, the spherical aberration 
phase~\cite{torok1995electromagnetic, torok1996electromagnetic, dutra2007polarization, dutra2014absolute}
\begin{equation}
\Phi_{\rm g-w}(\theta)=k_g\left( -L\cos\theta+NL_{c}\cos\alpha\right), \label{aberration_interface}
\end{equation}
 arises from 
refraction at this planar water-glass interface. 
 Here, $\theta=\arcsin(n_{\rm w}\sin\alpha/n_{g})$
is the refraction angle in the glass medium, $N=n_{\rm w}/n_{g}$ is the relative refractive index for the interface, $k_g=2\pi n_g/\lambda$ is the wavenumber for propagation in glass. The aberration function $\Phi_{\rm g-w}(\theta)$ scales with the lengths $L$ and $L_c$ representing the positions of the focal plane and of the microsphere center of mass, respectively, both relative to the water-glass interface. 

 After refraction, the scattered light is collected by the microscope objective and then propagates throughout the tube lens (focal length $f'$) of low numerical aperture, where it is eventually focused on the camera. The same process occurs for the field associated with the illumination that propagates towards the tube lens to ultimately interfere with the field scattered by the microsphere at the camera position. 
 The total electric field after propagation through the imaging system is given in Appendix B. 
 
The imaging setup shown in Fig.~\ref{fig1}(a) ensures that light is always detected in the forward direction. As a result, the conditions for observing optical rotatory power are fulfilled regardless of whether the polarization handedness is preserved during scattering—that is, independent of the scatterer's duality properties~\cite{fernandez2013necessary,fernandez2013electromagnetic,fernandez2015dual}. In the forward direction, rotatory power requires only the breaking of spatial inversion symmetry, which in our case is achieved by the presence of a chiral scatterer~\cite{fernandez2013necessary,fernandez2013electromagnetic,fernandez2015dual}. Therefore, our approach is consistent with the contemporary understanding of rotatory power that takes into account the role of the dual symmetry in electromagnetism~\cite{fernandez2013necessary,fernandez2013electromagnetic,fernandez2015dual}. 
In addition, we stress that our main focus lies at CD-like effects captured by the Stokes parameter $S_3$, and not at optical rotatory power, encoded in the Stokes parameter $S_2$.

We perform polarimetry of the detected total field using the  Stokes parameters $S_0$, $S_1$, $S_2$, and $S_3$ \cite{shurcliff1962polarized,bickel85,goldstein03,schaefer07,Cox23} written as functions of the total electric field components, namely: $S_0=E_x E_x^{*}+E_y E_y^{*}$ that represents the intensity of the detected total field; $S_1=E_x E_x^{*}-E_y E_y^{*}$ that gives the amount of horizontal and vertical linear polarizations; $S_2=E_x E_y^{*}+E_y E_x^{*}$ that describes the amount of diagonal polarizations along the $45^{\circ}$ and $135^{\circ}$ directions; and $S_3=i(E_x E_y^{*}-E_y E_x^{*})$ that accounts for the amount of circular polarization in the left and right directions. In the next section, we calculate $S_1$, $S_2$, and $S_3$ normalized by the parameter $S_0$.

\section{Results and discussion}

In the following, we consider realistic values for the optical system parameters: tube lens focal length $f'=20\,\rm cm$, and glass refractive index $n_g=1.51$. In most examples, we take 
$\rm NA=1.3$ for the objective lens and obtain its focal length $f$  from the typical magnification $M=100\times$ of the objective as $f=n_gf'/M=0.5\,{\rm cm}.$ \cite{gomez2021nonparaxial}  We account for the dispersion of water encoded in the refractive index formula $n_{\rm w}=1.3219+3.631\times 10^{-3}/\lambda[\mu{\rm m}]$ and  set $\epsilon=2.1$ for the relative electric permittivity of the microsphere \cite{ali2020enantioselective}. We also consider the microsphere centered on the optical axis and touching the water-glass interface ($L_c=a$) and take the focal plane at the water-glass interface  ($L=0$),
 to reduce the spherical aberration arising from refraction at this interface. 

Figure~\ref{fig1}(b) highlights one of the key findings of this work, namely the fact that the Stokes parameter $S_3$ not only can be non-vanishing but also may reach large values for lossless spherical particles 
particularly for large (non-paraxial) values of 
the Numerical Aperture (NA) of the objective employed in the proposed imaging setup depicted in Fig.~\ref{fig1}(a). $S_{3}$ usually describes the well-known effect of Circular Dichroism (CD), which is the differential absorption of left- and
right-handed circularly polarized light\cite{schnoering2018three}. To investigate how the non-paraxial 
nature of the optical system influences the detected polarization,  we analyze the
variation of the Stokes parameters with the 
objective NA in Fig.~\ref{fig1}(b). We choose the wavelength $\lambda=0.464\,\mu{\rm m}$ and consider 
a microsphere of radius $a=1.5\,\mu{\rm m}$ and chirality parameter $\kappa=-0.02.$
The detected polarization is approximately left-handed circular  $(\sigma=+1)$ with $S_3\approx +1$  when NA$~=\mbox{NA}_{\rm max}=1.3$. 
As one changes NA, we consider a fixed value for the radius of the objective back aperture  in order to collect the same power in all cases. As a consequence, the focal length changes according to 
 $f=f_{\rm max}\,\mbox{NA}_{\rm max}/\mbox{NA},$ where $f_{\rm max}=0.5\,{\rm cm}$ is the focal length for $\mbox{NA}_{\rm max}=1.3$ as mentioned earlier.
 
 In the paraxial limit, which corresponds to detecting a single forward plane wave (NA$\rightarrow 0$), the horizontal linear polarization of the incident field is approximately conserved during scattering, and therefore $S_1\rightarrow 1$ as shown in Fig.~\ref{fig1}(b). 
As the numerical aperture increases, the horizontal polarization rotates counterclockwise, passing through states close to maximum diagonal and vertical polarizations around ${\rm NA}\sim 0.1$ and ${\rm NA}\sim 0.15,$ respectively.
 For higher values of NA, while the degree of circular polarization, represented by the Stokes parameter $S_3$, 
 increases non-monotonically, the parameters $S_1$ and $S_2$ reduce in a non-trivial manner until they reach zero for ${\rm NA}_{\rm max}=1.3$, while the detected beam reaches an approximately pure state of circular polarization.
 
Overall, Fig.~\ref{fig1}(b) puts in evidence the importance of considering high NA and the corresponding large off-axis scattering angles in order to observe a nonvaninshing $S_3$ for dielectric chiral particles. Indeed for small NA, which is typically the case of standard CD spectrometers~\cite{adhikari2022optically}, not only $S_3$ is small but also it is smaller than $S_2$. Indeed,  rotatory power is the most appropriate way to probe chirality of dielectric chiral particles in the paraxial regime. In contrast, as one increases NA, $S_3$ significantly overcomes $S_2$ suggesting that in this case addressing $S_3$ should facilitate the characterization of the optical response of lossless chiral particles. 

It is also important to emphasize that our results are consistent with the conservation law for optical chirality~\cite{poulikakos2016optical}. Indeed, we find that the total field (incident + scattered) carries a net-zero chirality flux through a Gaussian spherical surface enclosing the Mie scatterer, provided that the host medium is non-absorbing (see Appendix C). This result is consistent with non-zero values of $S_{3}$ obtained in the considered detection geometry, in which scattered Fourier components are collected only within the forward scattering region delimited by the objective numerical aperture.
 
\begin{figure}[t!]
    \centering
 \includegraphics[scale=0.5]{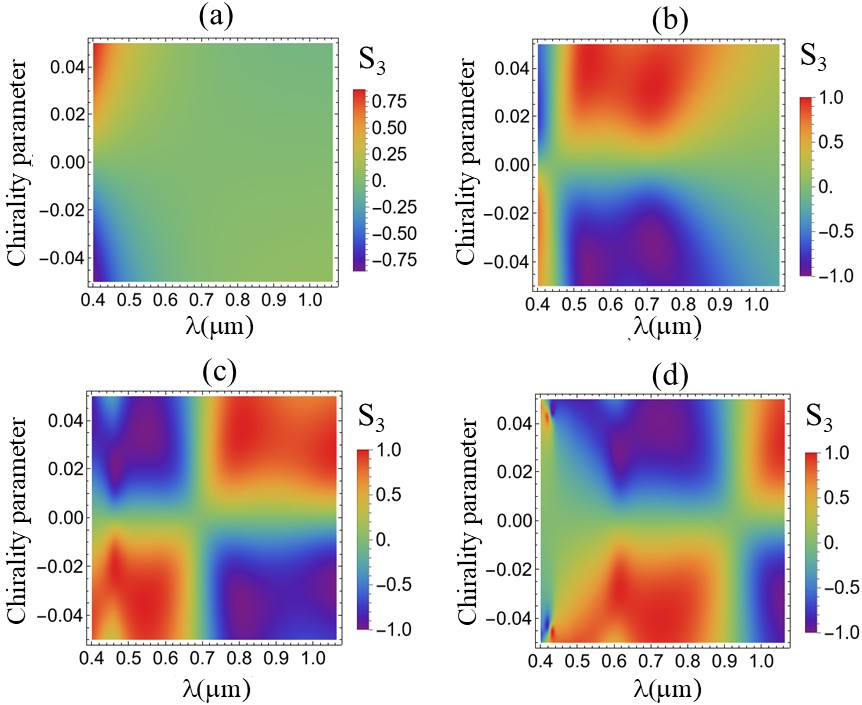}
 \vspace{-0.3cm}
   \caption{Color maps of the Stokes parameter $S_3$, normalized by the parameter $S_0$, as a function of the illumination wavelength $\lambda$ and the chirality parameter $\kappa$, for different values of microsphere radius (a) $a=0.5 \,\mu{\rm m}$, (b) $a=1.0\,\mu{\rm m}$, (c) $a=1.5\,\mu{\rm m}$ and (d) $a=2.0\,\mu{\rm m}$. The value of the numerical aperture is  NA $=1.3$.}
  \label{fig2} 
   \end{figure}

While the results shown in Fig.~\ref{fig1}(b) are valid for particular values of the chirality parameter $\kappa$ and of the incident wavelength $\lambda$, we explore in the color maps of $S_3$ shown 
in Fig.~\ref{fig2} the full parameter space defined by  $\kappa$  and $\lambda.$ 
The analysis of Fig.~\ref{fig2} confirms that Mie scattering by a lossless chiral particle can exhibit large values of $S_3$ over a broad range of values of $\kappa$ and for a wide range of visible frequencies, which mimics the CD effect. Our findings challenge this traditional scenario, showing that an analog to CD, associated to a nonvanishing value of $S_3$,  may also exist for lossless chiral particles provided the Mie regime is met. Indeed, Fig.~\ref{fig2} (a), in which $S_3$ is calculated for smallest sphere's radii $a$,  show that $S_3$ is negligible in the dipolar regime ($\lambda\gg a$). The right-hand-side of  Fig.~\ref{fig2} (b) also indicates that $S_3$ goes to zero for large wavelengths.  In the opposite limit of geometrical optics ($\lambda\ll a$), $S_3$ is also very small as shown in a more evident way in 
the left-hand-side (smallest values of $\lambda$)  of
Fig.~\ref{fig2}(d) , which corresponds to the largest value of $a$ shown in the figure. Altogether these results highlight the importance of addressing the Mie regime  ($\lambda\sim a$) in order to achieve large values of $S_3$ even for single, lossless chiral spherical particles, unveiling an effect that is the analog of CD for all-dielectric chiral particles.    
Remarkably, Fig.~\ref{fig2} also demonstrates that $S_3$ may change sign by varying the incident wavelength without changing the sign of $\kappa$. This effect does not occur in the dipolar regime (see Fig.~\ref{fig2}(a)) and emerges only in the Mie regime, as the majority of important results in this work. 

As a matter of consistency, in Fig.~\ref{fig2} the polarimetry of the total detected field in the focal plane of the tube lens shows that in the limit of an achiral microsphere ($\kappa \rightarrow 0$), $S_3 \rightarrow 0 $ and $S_1 \rightarrow 1$. In this case, the helicities $\sigma=\pm 1$ are not unbalanced during detection, and the total detected polarization is similar to the initial linear polarization $\mathbf{\hat{x}}$. For achiral microspheres, both helicities of the Fourier components, collected along the optical axis in the focal plane of the tube lens, are scattered with same amplitude $(a_j+b_j)$ (see Appendix B), thus reflecting the conservation of polarization state in this type of detection.
 In contrast, the chirality of the microsphere induces an unbalance of helicities during detection.  Indeed, according to Fig.~\ref{fig2}, the degree of circular polarization of the total detected field increases, and becomes fully circularly polarized when $S_{3}=\pm 1$, as the absolute value of the chirality parameter increases in certain wavelength ranges.

\begin{figure}[t!]
    \centering
 \includegraphics[scale=0.5]{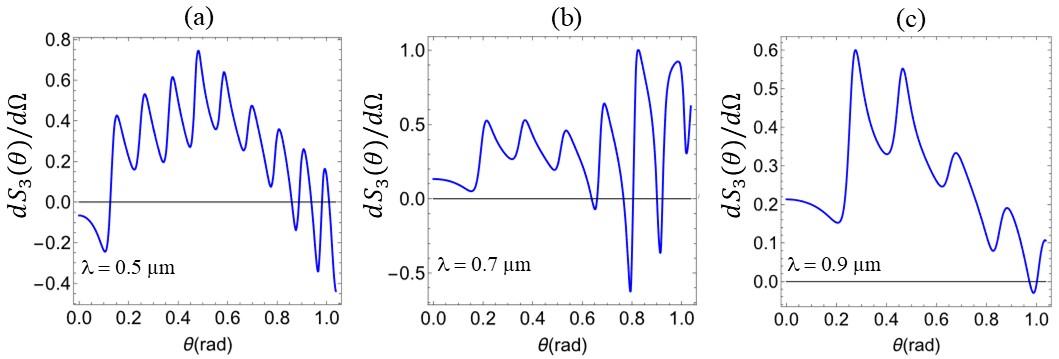}
 \vspace{-0.3cm}
   \caption{
   Differential Stokes parameter $\frac{dS_3}{d\Omega}$ (normalized by $S_0$), associated with the field of an individual conical shell of scattered plane waves superimposed to the illumination field, as a function of the scattering angle $\theta$ for different wavelengths: (a) $\lambda=0.5\,\mu{\rm m}$, (b) $\lambda=0.7\,\mu{\rm m}$ and (c) $\lambda=0.9\, \mu {\rm m}$. We consider a microsphere of radius $a=1.5\, \mu{\rm m}$ and chirality parameter $\kappa=-0.02$.}
  \label{fig3} 
   \end{figure}

It is important to emphasize that the calculation of the Stokes parameters in our detection geometry involves the coherent superposition of the Fourier components that are scattered in different directions, and that interfere coherently not only with the incident field but also between themselves. In Figure~\ref{fig3} we show the differential Stokes parameter $ \frac{dS_{3}}{d \Omega} (\theta)$ ($d \Omega$ being the solid angle of a thin conical angular shell) associated with the field of a single, individual scattered conical shell of plane waves superimposed with the incident field as a function of the scattering angle $\theta$. 
For example note that in Fig.~\ref{fig3}(c) $ \frac{dS_{3}}{d \Omega} (\theta)$ is positive for the vast majority of values of $\theta$ so that its integral over $\theta$ must be clearly positive as well. In contrast, the actual value of $S_{3}$ is negative, as shown in Fig.~\ref{fig2}(c) at $\lambda =0.9 \, \mu$m and for the parameters corresponding to Fig.~\ref{fig3}(c). Indeed, the integral of $ \frac{dS_{3}}{d \Omega} (\theta)$ represents an incoherent sum of intensities emerging from the polarimeter whereas our calculations are based on a coherent superposition of all scattered single wave components. As a result, altogether these findings show the crucial role of the coherent superposition and interference of different scattered plane wave components in the sign change of the Stokes parameter $S_3$ shown in Fig.~\ref{fig2}(c).

\begin{figure}[t!]
    \centering
 \includegraphics[scale=0.35]{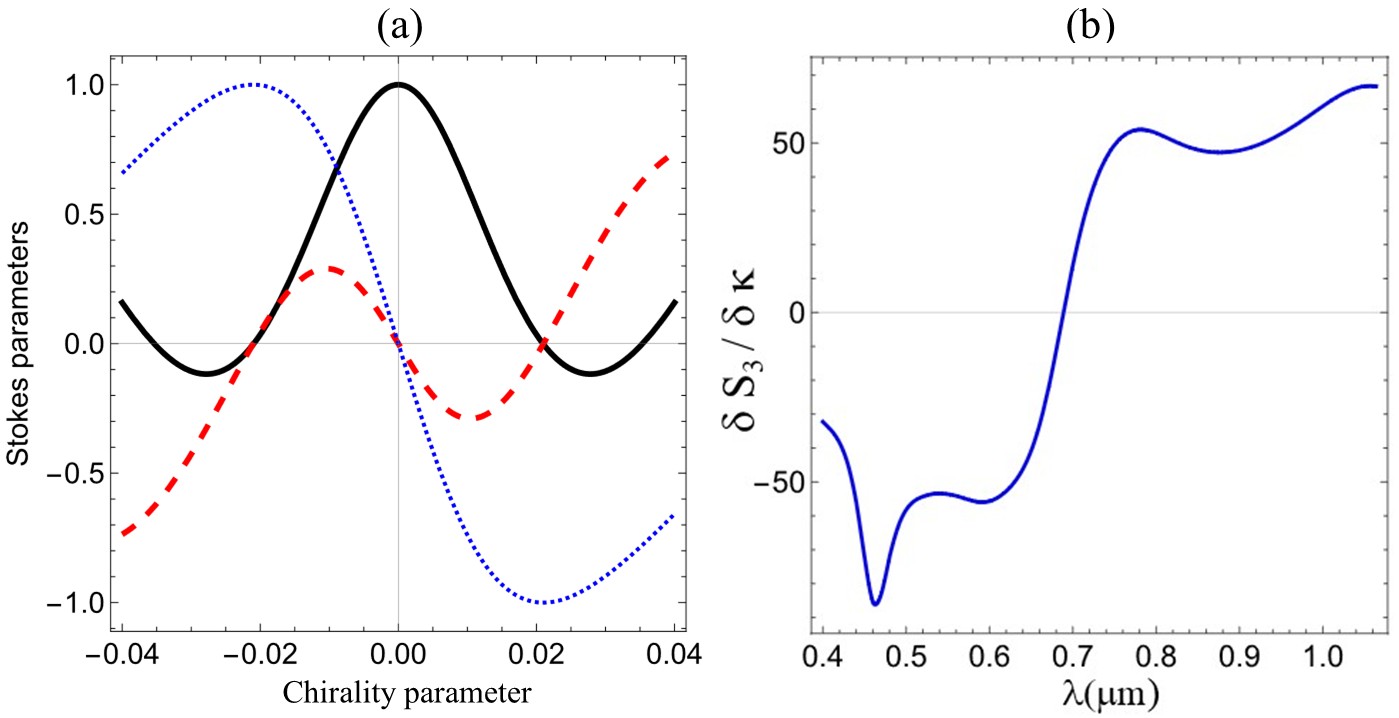}
 \vspace{-0.3cm}
   \caption{(a) Stokes parameters $S_1$ (black solid line), $S_2$ (red dashed line), and $S_3$ (blue dotted line), normalized by the Stokes parameter $S_0$, as functions of the microsphere chiral parameter $\kappa$ for the wavelength
$\lambda=0.464\,\mu{\rm m}.$
    (b) Slope $\delta S_3/ \delta \kappa$ near $\kappa=0$ as a function of $\lambda.$ The  microsphere radius is  $1.5\,\mu{\rm m}$ and the numerical aperture is NA$~=1.3$ for both panels.}
  \label{fig4} 
  \end{figure}

  For the value of the microsphere radius $a = 1.5\,\mu{\rm m}$ and  wavelength $\lambda=0.464\,\mu{\rm m}$, we study in Fig.~\ref{fig4} the dependence of the Stokes parameters on the chirality parameter $\kappa$.  
  Interestingly, in this case $\vert S_{3} \vert$ is  larger than $ \vert S_{2} \vert$, even for very small values of $\kappa$. This result shows that for a dielectric particle with chirality parameters of the order of naturally occurring materials, the CD-like effect encoded in $S_{3}$ may overcome the optical rotatory power, related to $S_{2}$. Figure~\ref{fig4}(a) also shows that the detected polarization evolves from a state of horizontal linear polarization ($S_1=1$), for an achiral microsphere ($\kappa=0$), until it reaches maximum circular polarization states with $S_3=\pm 1$, for chirality parameters close to $\kappa\approx\mp 0.02$, respectively. 
  Figure~\ref{fig4}(a) reveals that in the vicinities of $\kappa=0$, $S_3$ exhibits a linear dependence on $\kappa$, which allows one to estimate the sensitivity of the Stokes parameter $S_3$ required to determine  small chirality parameters. Indeed, Fig.~\ref{fig4}(b) shows that $ \vert \delta S_{3}/ \delta\kappa \vert \approx 10^{2}$ so that, considering that the typical sensitivity of state-of-the-art CD spectrometers is of the order $ S_{3}/S_{0} \approx 10^{-3}$~\cite{li2023strong}, one could detect chirality parameters as small as  $\vert \delta\kappa \vert \approx 10^{-5}$. This result opens up the possibility of characterizing the chirality parameter of isolated dielectric particles that are embedded in, doped with, or attached to naturally occurring chiral materials~\cite{mcarthur2025observation}.

\begin{figure}[t!]
    \centering
 \includegraphics[scale=0.3]{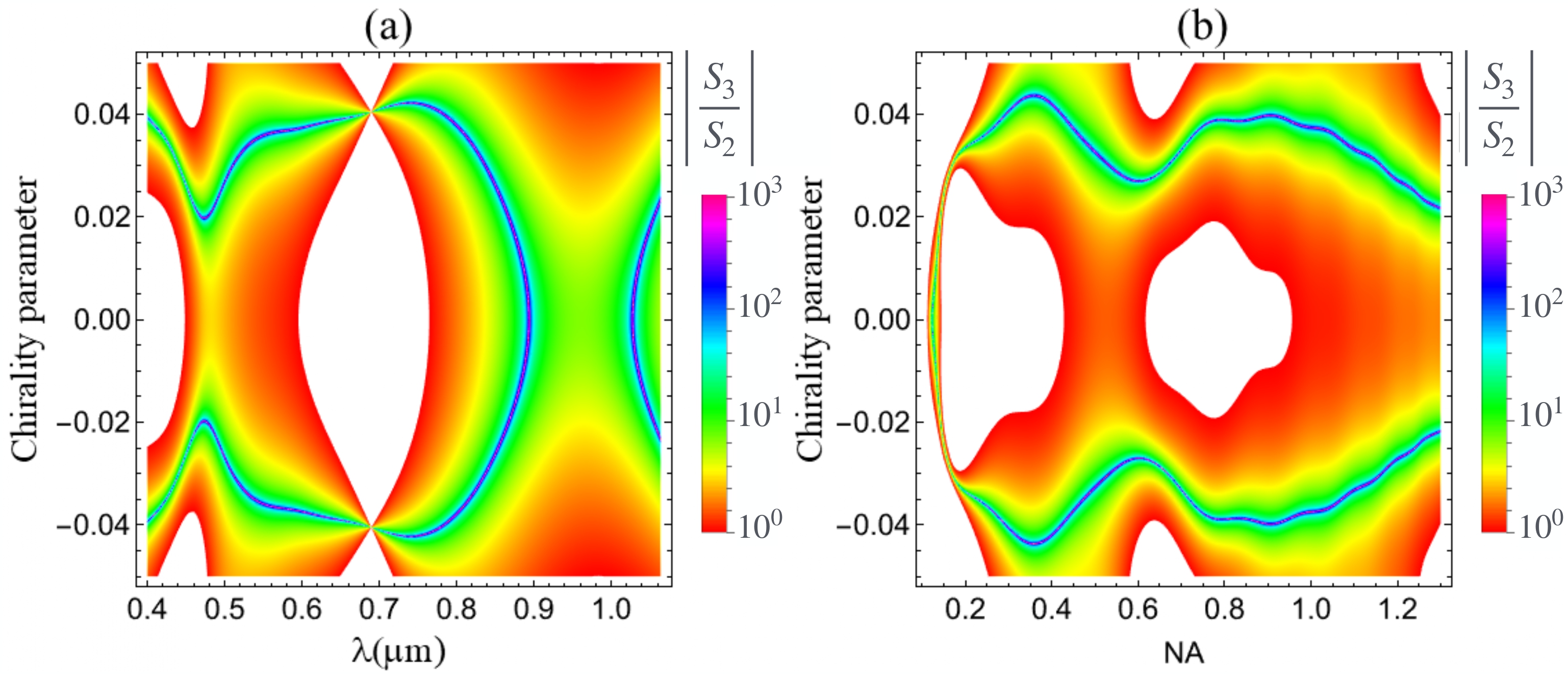}
 \vspace{-0.3cm}
   \caption{ Color maps of $|S_3/S_2|$ (log scale) as function of
   the chirality parameter $\kappa$ and 
   of (a) wavelength $\lambda$ or (b)  numerical aperture NA. 
   We take NA$~=1.3$ for the former and 
   $\lambda=0.464\,\mu{\rm m}$ for the latter. 
   The microsphere radius is $1.5\,\mu{\rm m}$ and the regions where $|S_3/S_2|<1$ are white. }
  \label{fig5} 
  \end{figure}

Besides demonstrating detectable values of the Stokes parameter $S_3$ from the light scattered by chiral lossless spheres, it is important to compare these values to the Stokes parameter $S_2$, which gives optical rotatory power. At first glance, one could argue that $S_2$ should always dominate over $S_3$ due to the fact the particle is lossless, regardless of the detection setup, wavelength $\lambda$ and the value of NA. However Fig.~\ref{fig5}, where the ratio $\vert S_3/S_2 \vert$ is calculated as a function of $\kappa$ and $\lambda$ (panel (a)) and of $\kappa$ and NA (panel (b)), demonstrates that this is not true. In fact, $\vert S_3 \vert$ can be ten times larger than $\vert S_2 \vert$ for a broad, nonresonant range of wavelengths and values of NA. In contrast, with single chiral plasmonic particles large values of CD are typically achieved in a narrow frequency window due to a plasmon resonance, which is unavoidably associated to detrimental losses\cite{hentschel17,garcia2019enhanced,sun23}. For lossless spheres $S_3$ can be even two orders of magnitude higher than $S_2$ at specific Mie resonances with high quality factors, which do not imply losses. Figures~\ref{fig5}(a) and \ref{fig5}(b) corroborates the previous results that disclose the conditions for the existence of a sizable CD-like effect for lossless chiral spheres, namely the Mie regime ($a \simeq \lambda$) and large NA, respectively. The white regions in Fig.~\ref{fig5} correspond to scenarios where $S_2$ dominates over $S_3$. In such cases, optical rotatory power is expected to be a more suitable metric for characterizing the chiroptical response of a single lossless chiral particle. Ideally, Fig.~\ref{fig5} serves as a theoretical roadmap to facilitate more efficient enantioselection and chiral characterization of lossless, isolated chiral spheres—whether this is achieved through analysis of optical rotatory power or a CD-like signal.


Finally, it is important to emphasize that a non-vanishing value of $S_3$ is not related to any optical anisotropy of the system since the scattering sphere is homogeneous and isotropic. As a result, the Mueller matrix~\cite{bickel85} describing the scattered radiation would capture only genuine CD-like terms.

\section{Conclusion}

In conclusion, we unveil an alternative chiroptical response of all-dielectric Mie chiral particles that is phenomenologically analogous to circular dichroism, well-known in absorbing media. This phenomenon shows up as large values of the Stokes parameter $S_{3}$ for a broad frequency range that we demonstrate to only exist in the Mie scattering regime and for large numerical apertures, an experimentally feasible scenario that nevertheless is not the typical configuration of standard spectrometers, which often use large off-axis detection. By disclosing that chiral Mie particles exhibit an effect that mimics circular dichroism, we pave the way for polarimetric applications in Mie resonant metaphotonics (also known as Mie-tronics), where all-dielectric scattering particles substitute traditional plasmonic structures to achieve many practical applications for subwavelength trapping
of light~\cite{kivshar22}. Our results provide then a link between chiral photonics of single particles and Mie-tronics, enabling potential applications that involve directional scattering with the generation of pure circular polarization states ($S_3 = \pm 1$), corresponding to maximal spin angular momentum transfer, enantioselection and characterization of the chiroptical response of isolated chiral, lossless particles.


\section{Acknowledgments}
We thank Arthur L. Fonseca and Kain\~a Diniz for discussions. 
This work was supported by the CAPES-COFECUB collaboration project n° Ph1030/24 and is also part of the Interdisciplinary Thematic Institute QMat of the University of Strasbourg, CNRS and Inserm, supported by the following programs: IdEx Unistra (ANR-10-IDEX- 0002), SFRI STRATUS project (ANR-20-SFRI-0012), CSC (ANR-10-LABX- 0026, project ITI-CSC-CGE-22) and USIAS (ANR-10-IDEX- 0002-02). The support of the CNRS "Risky and High-Impact Research" program (ANR-24-RRII-0001, project POLARITONIC) is also acknowledged. The authors also thank
Conselho Nacional de Desenvolvimento
Científico e Tecnológico (CNPq–Brazil), Coordenação
de Aperfeiçamento de Pessoal de Nível Superior
(CAPES–Brazil), Instituto Nacional de Ciência e Tecnologia
de Fluidos Complexos (INCT-FCx), and the Research
Foundations of the States of Rio de Janeiro (FAPERJ) and
São Paulo (FAPESP)
for finantial support. 

\appendix

\section{Mie Coefficients for a Chiral Sphere}

By applying the boundary conditions on the surface of the microsphere of radius $a$, effective scattering coefficients $A_j^{\sigma}$ and $B_j^{\sigma}$ associated with the electric and magnetic multipoles are obtained, respectively, and expressed by

\begin{equation}
A_j^{\sigma} = a_j + i\sigma d_j \,\,\,\,\,\, and \quad
B_j^{\sigma} = b_j - i\sigma c_j,
\end{equation}
in which the scattering coefficients $a_j$, $b_j$, $c_j$, and $d_j$ are written in terms of the size parameter $x=\omega a/c$ through the following expressions \cite{bohren1974light}:

\begin{align}
a_j(x) &=  \Delta^{-1}_j(x) \left[V_j^{R}(x) A_j^{L}(x) + V_j^{L}(x) A_j^{R}(x)\right] \\
b_j(x) &=  \Delta^{-1}_j(x) \left[W_j^{R}(x) B_j^{L}(x) + W_j^{L}(x) B_j^{R}(x) \right] \\
c_j(x) &= i\Delta^{-1}_j(x) \left[W_j^{R}(x) A_j^{L}(x) - W_j^{L}(x) A_j^{R}(x)\right] 
\end{align}
in which we use the following auxiliary functions
\begin{equation}
\Delta_j(x) = W_j^{L}(x) V_j^{R} (x) + W_j^{R} (x) V_j^{L}(x)
\end{equation}
\begin{align}
W_j^{L,R} (x) &= M\psi_j(N_{L,R}x) \xi'_j(x) - \xi_j(x)\psi'_j(N_{L,R} x) \\
V_j^{L,R}(x) &= \psi_j(N_{L,R} x) \xi'_j(x) - M \xi_j(x)\psi'_j(N_{L,R} x) \\
A_j^{L,R}(x) &= M\psi_j(N_{L,R} x) \psi'_j(x) - \psi_j(x)\psi'_j(N_{L,R} x) \\
B_j^{L,R}(x) &= \psi_j(N_{L,R} x) \psi'_j(x) - M\psi_j(x)\psi'_j(N_{L,R} x)\,. 
\end{align}
with the relative refractive index \(N_{L,R} = (\sqrt{\epsilon \mu} \pm \kappa)/n_\text{w}\) and the relative impedance \(M = \sqrt{\mu/\epsilon} \, n_\text{w}\). The scattering coefficients are represented using the Riccati-Bessel functions \(\psi_j(z) = z j_j(z)\) and \(\xi_j(z) = z h_j^{(1)}(z)\). The mixed polarization scattering coefficients satisfy \(c_j = -d_j\). In the limit where the chirality parameter approaches zero, the scattering coefficients simplify to the usual Mie coefficients \(a_j = A_j/W_j\), \(b_j = B_j/V_j\), and \(c_j = d_j = 0\).

\section{Total Field Detected}

In the situation where detection is carried out in the focal plane of the tube lens, with $\rho_t=0$ and $z_t=0$, we have the following expression for the detected scattered field in the forward direction:
\begin{equation}
\mathbf{E}_{s,tube}=\frac{E_{0}}{4}\frac{f}{n_gf'}e^{ik_g f} e^{ik_{0}(D+f')}\sum_{j=1}^{\infty}\sum_{\sigma=-1}^{+1}(2j+1)(A_j^{\sigma}+B_j^{\sigma})
\int_{0}^{\theta_0}\,d^{j}_{1,1}(\alpha)f(\theta)d\theta\,(\mathbf{\hat{x}}+i\sigma\mathbf{\hat{y}}),
\end{equation} 
where $f(\theta)=(\sin\theta/\cos\alpha)(\cos\theta)^{3/2} \,T_{\perp}(\theta)e^{i\Phi_{g-w}(\theta)}$, $D$ is the distance between the objective and the tube lens, and $k_0=2\pi/\lambda$ the wave vector in air.  Here, $T_{\perp}(\theta)$ is the Fresnel refraction amplitude for the water-glass interface. 

For an achiral sphere $\kappa=0$, $A_{j}^{\sigma}=a_j$ and $B_{j}^{\sigma}=b_j$, and both  helicities are scattered with the same weights, that is, with the same  amplitude $(a_j+b_j)$, such that the expression for the detected scattered field  is simplified, conserving the incident linear polarization $\hat{\mathbf{x}}$:
\begin{equation}
\mathbf{E}_{s,tube}^{achiral}=\frac{E_{0}}{2}\frac{f}{n_gf'}e^{ik_g f} e^{ik_{0}(D+f')}\sum_{j=1}^{\infty}(2j+1)(a_j+b_j)
\int_{0}^{\theta_0}\,d^{j}_{1,1}(\alpha)f(\theta)d\theta\,\mathbf{\hat{x}}.
\end{equation} 
In turn, the illumination field propagated through the tube lens is expressed by

\begin{equation}
\mathbf{E}_{\rm in,tube}(\mathbf{r}_t)=
-\frac{E_{0}f}{f'}\frac{2n_{\rm w}}{n_{\rm w}+n_g}e^{i(k_{\rm w}L_{c}-k_{g}L_{g})}e^{ik_{0}(f'+D)}e^{ik_gf}\,\mathbf{\hat{x}}. \label{fieldintube}
\end{equation}

\section{Optical chirality Flux in the Mie Scattering Scenario}

In this section, we verify the conservation law for the optical chirality flux \cite{poulikakos2016optical} in the scenario of scattering by a Mie sphere immersed in a homogeneous medium.

\subsection{Chirality Flux}

Using Maxwell's equations in the absence of sources, the optical chirality flux density can be expressed as 

\begin{equation}
\mathbf{F}=\frac{i\omega}{4}(\varepsilon^{*}\mathbf{E}\times\mathbf{E}^{*}-\mu\,\mathbf{H}^{*}\times\mathbf{H}). \label{eq1}
\end{equation}

We evaluate the time-averaged chirality flux $\Phi$ over an imaginary sphere of infinite radius, enclosing the Mie sphere, in terms of the total fields given by the superposition of incident and scattered fields $\mathbf{E}_{tot}=\mathbf{E}_{in}+\mathbf{E}_{s}$ and $\mathbf{H}_{tot}=\mathbf{H}_{in}+\mathbf{H}_{s}$:

\begin{equation}
\Phi= \Re\biggl(\oint_{s}\,\mathbf{F}\cdot\hat{r}\,dS\biggr). \label{eq2}
\end{equation}

It is convenient to expand the total electromagnetic field ($\mathbf{E}_{tot},\mathbf{H}_{tot}$) in spherical waves using the electric (E) and magnetic (M) multipoles, $\mathbf{E}_{tot}=ik(\, \mathbf{I}^{E}_{z}-(\mu\,c/n_{1})\mathbf{I}^{M}_{x}\,)$ and $\mathbf{H}_{tot}=ik(\, \mathbf{I}^{M}_{z}+(\varepsilon\,c/n_{1})\mathbf{I}^{E}_{x}\,)$:

\begin{equation}
\mathbf{I}^{E,M}_{z}=i\frac{ C^{E,M}}{kr}\sum_{JM}\Gamma^{E,M}_{J,M}(kr)(i)\biggl(\,\hat{r}\times\mathbf{L}(\,Y_{J,M}(\theta,\phi)\,)\,\biggr) \label{eq3}
\end{equation}

\begin{equation}
\mathbf{I}^{E,M}_{x}= C^{E,M}\sum_{JM}\Omega^{E,M}_{J,M}(kr)(i)\mathbf{L}(\,Y_{J,M}(\theta,\phi)\,),
\end{equation}
where $\mathbf{L}=-i{\bf r}\times \boldsymbol{\nabla}.$
We keep only the tangential components of the fields, since the radial components do not contribute to the flux in the radiation zone.
The scattered amplitudes and the radial dependence are taken into account implicitly in functions $\Gamma^{E,M}_{J,M}(kr)$ and $\Omega^{E,M}_{J,M}(kr)$. Substituting eq. \ref{eq1} into \ref{eq2}, together with the fields, we obtain:

\[
\Phi=-\frac{\omega\,k^2}{4}\Im\biggl[\oint_{s}\,\biggl(\varepsilon^{*}(\mathbf{I}_{z}^{E}\times\mathbf{I}_{z}^{E*})\cdot\hat{r}-\mu\,(\mathbf{I}_{z}^{M*}\times\mathbf{I}_{z}^{M})\cdot\hat{r}\biggr)\,dS\biggr]+\]\[-\frac{\omega\,k^2}{4}\Im\biggl[\oint_{s}\,\biggl(\frac{\varepsilon^{*}\mu^2c^2}{n_1^{2}}(\mathbf{I}_{x}^{M}\times\mathbf{I}_{x}^{M*})\cdot\hat{r}-\frac{\mu\,\varepsilon^2c^2}{n_1^{2}}(\mathbf{I}_{x}^{E*}\times\mathbf{I}_{x}^{E})\cdot\hat{r}\biggr)\,dS\biggr]\]\begin{equation}+
\frac{\omega\,k^2}{2}\Im\biggl[\oint_{s}\,\biggl(\varepsilon^{*}\Re\biggl(\frac{\mu^{*}c}{n_1^{*}}(\mathbf{I}_{z}^{E}\times\mathbf{I}_{x}^{M*})\cdot\hat{r}\biggr)-\mu\,\Re\biggl(\frac{\varepsilon c}{n_1}(\mathbf{I}_{z}^{M*}\times\mathbf{I}_{x}^{E})\cdot\hat{r}\biggr)\biggr)\,dS\biggr]    \label{eq5}
\end{equation}

According to equation \ref{eq5}, two types of terms, quadratic and cross terms in the multipoles, contribute to the chirality flux.

\subsection{Quadratic Terms}

The first two terms in equation \ref{eq5} involve surface integrals of the type $I=\oint_{s} (\mathbf{I}_{z}^{E,M}\times\mathbf{I}_{z}^{E,M*})\cdot\hat{r}\,dS$. Using equation \ref{eq3}, we obtain

\begin{equation}
I=-\frac{C^{{E,M}^2}}{k^2r^2}\sum_{JM}\sum_{J'M'}\Gamma^{E,M}_{JM}\Gamma^{E,M*}_{J'M'}\oint_{s}
\biggl(\,\hat{r}\times\mathbf{L}(\,Y_{J,M}(\theta,\phi)\,)\,\biggr)\times \biggl(\,\hat{r}\times\mathbf{L}(\,Y_{J',M'}^{*}(\theta,\phi)\,)\,\biggr)\cdot\hat{r}dS.
\end{equation}

Using $dS=r^2sin\theta d\theta d\phi$, $\mathbf{L}=-(i)\biggl(\hat{\phi}\,\partial_{\theta}Y(\theta,\phi)-\hat{\theta}\,\partial_{\phi}Y(\theta,\phi)/\sin\theta \biggr)$, $\partial_{\phi} Y_{J,M}(\theta,\phi)=iM Y_{J,M}(\theta,\phi)$ and $\int_0^{2\pi}e^{i(M'-M)\phi}\,d\phi=2\pi\delta_{M,M'}$, we find that

\begin{equation}
I=2\pi\,ir^2\frac{C^{{E,M}^2}}{k^2r^2}\sum_{JM}\sum_{J'}\Gamma^{E,M}_{JM}\Gamma^{E,M*}_{J'M} M 
\int_{0}^{\pi}\frac{\partial}{\partial \theta}\biggl( Y_{J,M}(\theta,0) Y^{*}_{J',M}(\theta,0)\biggr)\,d\theta = 0
\end{equation}

Similarly, it can be shown that the quadratic terms $\oint_{s}(\mathbf{I}_{x}^{E,M}\times\mathbf{I}_{x}^{E,M*})\cdot\hat{r}\,dS$ also do not contribute to the flux. For a homogeneous medium, the quadratic terms in the electric and magnetic multipoles do not contribute to the chirality flux.

\subsection{Cross Terms}

The contribution to the chirality flux comes only from the cross terms in the multipoles (E) and (M) in the situation of a homogeneous host medium with absorption:

\begin{equation}
\Phi=
\frac{\omega\,k^2}{2}\biggl[\oint_{s}\,\biggl(\Im(\varepsilon^{*})\Re\biggl(\frac{\mu^{*}c}{n_1^{*}}(\mathbf{I}_{z}^{E}\times\mathbf{I}_{x}^{M*})\cdot\hat{r}\biggr)-\Im(\mu)\,\Re\biggl(\frac{\varepsilon c}{n_1}(\mathbf{I}_{z}^{M*}\times\mathbf{I}_{x}^{E})\cdot\hat{r}\biggr)\biggr)\,dS\biggr].    
\end{equation}

For a host medium without absorption, $\Im(\varepsilon)=\Im(\mu)=0$, and the chirality flux through the sphere vanishes in agreement with the general result of Ref.~\cite{poulikakos2016optical}.

\bibliographystyle{ieeetr}
\bibliography{main}

\end{document}